\documentclass[aps,prb,twocolumn,superscriptaddress]{revtex4-2} 

\usepackage[]{graphicx}
\usepackage[dvipsnames]{xcolor}
\usepackage[colorlinks=true,citecolor=NavyBlue,linkcolor=RubineRed,urlcolor=Cerulean,pdfencoding=auto]{hyperref}
\usepackage[]{amsmath}
\usepackage[]{amssymb}

\def\bra#1{\mathinner{\langle{#1}|}}
\def\ket#1{\mathinner{|{#1}\rangle}}
\def\braket#1{\mathinner{\langle{#1}\rangle}}

\def\abs#1{\mathinner{|{#1}|}}

\def\proj#1{\ket{#1}\bra{#1}}
\def\op#1#2{\ket{#1}\bra{#2}}

\def\comm#1#2{\mathinner{[{#1},{#2}]}}

\begin{document}


\title{Nuclear Spin Quantum 
Memory in Silicon Carbide}

\author{Benedikt Tissot}
\email[]{benedikt.tissot@uni-konstanz.de}
\affiliation{Department of Physics, University of Konstanz, D-78457 Konstanz, Germany}

\author{Michael Trupke}
\affiliation{Institute for Quantum Optics and Quantum Information (IQOQI) Vienna,
Austrian Academy of Sciences, Boltzmanngasse 3, 1090 Vienna, Austria}
\affiliation{Faculty of Physics, University of Vienna, Boltzmanngasse 5, 1090 Vienna, Austria}
\author{Philipp Koller}
\author{Thomas Astner}
\affiliation{Faculty of Physics, University of Vienna, Boltzmanngasse 5, 1090 Vienna, Austria}

\author{Guido Burkard}
\email[]{guido.burkard@uni-konstanz.de}
\affiliation{Department of Physics, University of Konstanz, D-78457 Konstanz, Germany}


\begin{abstract}
  Transition metal (TM) defects in silicon carbide (SiC) are a promising platform for applications in quantum technology.
  Some TM defects, e.g. vanadium, emit in one of the telecom bands, but the large ground state hyperfine manifold poses a problem for applications which require pure quantum states.
  We develop a driven, dissipative protocol to polarize the nuclear spin, based on a rigorous theoretical model of the defect.
  We further show that nuclear-spin polarization enables the use of well-known methods for initialization and long-time coherent storage of quantum states.
  The proposed nuclear-spin preparation protocol thus marks the first step towards an all-optically controlled integrated platform for quantum technology with TM defects in SiC.
\end{abstract}

\maketitle


\section{Introduction}

In the so called ``information age'' secure communication is becoming increasingly important.
Quantum communication is a viable option to achieve secure communication
via the protection of quantum channels by virtue of the no-cloning theorem.
In order to build scalable, real-world quantum networks, more progress in the domain of related quantum technologies, such as quantum memories, emitters, and many more~\cite{kimble08,aharonovich16,heshami16,awschalom21} needs to be made.
The fundamental problem to overcome is the ability to coherently control and selectively couple quantum systems,
while simultaneously isolating them from unwanted noise.

A much studied, promising system for quantum technology is the nitrogen vacancy center in diamond including neighboring spins \cite{he93,gaebel06,childress06,santori06,gali08,felton09,maze11,fuchs11,togan11,yale13,golter13,busaite20,hegde20}
(\cite{doherty13,suter17,pezzagna21} for reviews). While this system has a long coherence lifetime and has been used to demonstrate entanglement over more than one kilometer, its optical transition in the visible domain poses challenges for integration into photonic devices and requires wavelength conversion, which adds noise and leads to losses, for long-distance quantum communication \cite{Hensen2015,Janitz2020,Dreau2018}. 
%

The transition metal (TM) defects in silicon carbide (SiC) constitute a distinct but similarly promising class of defects.
These defect centers benefit from their host material which is well established in the semiconductor industry,
and from the availability of accessible transitions in the telecommunication bands~\cite{kaufmann97,baur97,bosma18,spindlberger19,gilardoni20,wolfowicz20,csore20}.
Recent experiments showed promising characteristics for the control of the nuclear spin of vanadium (V) defects in SiC which has optical transitions in the telecom O-band, for which high-performance photonic devices are available and long-distance quantum communication has been demonstrated over installed optical fiber links \cite{wolfowicz20}.

Building upon the current experiments~\cite{bosma18,spindlberger19,gilardoni20,wolfowicz20} and numerical calculations~\cite{csore20,gilardoni21}, as well as using the theoretical framework we developed in previous works~\cite{tissot21a,tissot21b},
in this article we identify promising qubit subsystems in TM defects and develop procedures to initialize these defects by nuclear spin polarization.
The initial polarization of the nuclear spin is a prerequisite for gaining control over a selected subsystem of levels (e.g., two levels for a qubit) from the multitude of nuclear spin states.

To optically pump the nuclear-spin polarization, we propose to use ratchet-type sequential population trapping into a polarized state.
The proposed method shows parallels to coherent population trapping into a dark state 
which is well established over a wide range of materials
from atoms\ \cite{gray78},
electrons in quantum dots\ \cite{xu08},
superconducting artificial atoms\ \cite{kelly10},
optomechanical systems\ \cite{dong12},
as well as NV centers in diamond\ \cite{santori06,togan11,yale13,golter13}.
A similar optical pumping was recently used to polarize Erbium nuclear spins 
\cite{ranvcic17,stuart21}.

This paper is organized as follows.
We introduce the physical model in Sec.~\ref{sec:phys}
and discuss possible qubit candidates in Sec.~\ref{sec:qubits}.
We then propose a protocol to polarize the nuclear spin in Sec.~\ref{sec:pol}, enabling the initialization of the system.
Next, we briefly discuss the prospect to engineer different protols based on technical limitations~\ref{sec:engineering} as well as a measurement of the polarization success~\ref{sec:measurement}.
Finally, we draw our conclusions in Sec.~\ref{sec:conclusion}.


\begin{figure*}
\centering
\includegraphics[width=18cm]{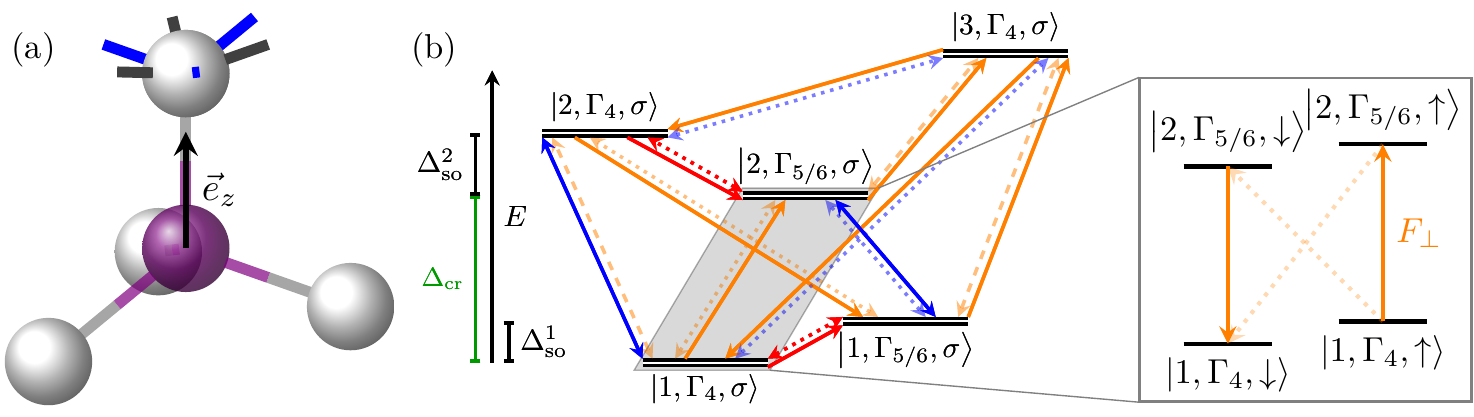} 
\caption{\label{fig:selection}
  Structure and energy level diagram of TM defects in SiC.
  (a) Schematic of a TM defect (purple) substituting a Si atom in SiC, with its tetragonally-arranged nearest-neighbor C atoms (grey).
  (b) Energy level diagram including selection rules for the electronic structure of a TM defect in SiC.
  The dependence of the transitions on the field direction is encoded in the line style in combination with the direction of the arrow.
  The {\color{blue}blue} lines denote allowed transitions for parallel fields \(B_z\) and \(\mathcal{E}_z\),
  while the {\color{orange}orange} lines are allowed for perpendicular magnetic  and electric fields, \(F_{\perp} = F_x + \mathrm{i} F_y\) (\(F = \mathcal{E},B\)).
  The selection rules for \(F_{\perp}^{*}\) enable the reverse process, in agreement with angular momentum conservation.
  Faded lines imply that the transition is only allowed due to spin-orbit mixing.
  The {\color{red}red} solid (dotted) lines are also for perpendicular fields, but allowed in the leading order for \(\mathcal{E}_{\perp}\) (\(B_{\perp}\)) and due to spin-orbit mixing of the states \(B_{\perp}\) (\(\mathcal{E}_{\perp}\)).
  The line styles furthermore imply the pseudospin selection rules:
  solid lines are pseudospin conserving transitions,
  where for the orange and red lines the direction of the arrow is relevant and implies the matrix elements according to \(\uparrow \Rightarrow \uparrow\) and \(\downarrow \Leftarrow \downarrow\).
  The blue dotted lines correspond to the spin flip transition.
  The orange and red dashed (dotted) correspond to the pseudospin \(\downarrow \Rightarrow \uparrow\) (\(\uparrow \Rightarrow \downarrow\)) transitions.
}
\end{figure*}

\section{Physical Model\label{sec:phys}}
The TM defects which we will focus on in this paper consist of a 
positively charged molybdenum (Mo$^{5+}$) or a neutral vanadium (V$^{4+}$) atom substituting a Si atom in 4H- or 6H-SiC.
Both of these defect atoms comprise one active electron in their \(D\)-shell~\cite{bosma18,spindlberger19,gilardoni20,csore20,wolfowicz20}.
In the presence of the surrounding crystal structure, the defects remain invariant under the transformations of the \(C_{3v}\) point group.
The symmetry reduction due to the crystal potential splits the \(D\)-shell into one orbital singlet and two orbital doublets.
Due to the spin-orbit interaction the electronic structure for the active electron is given by five Kramers doublets (KDs), which are pairs of states related to each other by time inversion.
We show the nearest neighbor structure of the defect in Fig.~\ref{fig:selection}(a) and the resulting energy level structure in Fig.~\ref{fig:selection}(b).

We concentrate on the interaction of the  active electron of the defect with the nuclear spin of the TM atom \cite{tissot21b},
which for the main V isotope is \(I = 7/2\) (abundance \(>99\%\)), while
\(I = 5/2\) for about \(25\%\) of the stable isotopes of Mo and $I=0$ for the remaining isotopes~\cite{audi03,meija16}.
In the following, we mainly focus on the V$^{4+}$ \(\alpha\) defect in 4H-SiC, though we note that the underlying theory is equally applicable to the other defects.
Polarization protocols and suitable qubit subsystems in other configurations can be derived analogously with the appropriately adjusted model parameters.

For simplicity we neglect the nuclear quadrupole interaction as well as the hyperfine interaction between different KDs, because both are expected to be small and were indeed not observed in recent experiments~\cite{wolfowicz20,tissot21b}.
We also neglect matrix elements between the KDs due to static fields, as they are suppressed by the large spin-orbit \(\Delta^i_{\mathrm{so}}\) or crystal-field splitting \(\Delta_{\mathrm{cr}}\) for magnetic fields \(\abs{\vec{B}}\ll\min_{i}|\Delta^i_{\mathrm{so}}/2\mu_{B}|\approx 6.5\,\)T for the V \(\alpha\) defect in 4H-SiC~\cite{wolfowicz20}.
Using these approximations we arrive at a block diagonal static Hamiltonian.
The blocks that describe the different KDs have the form
\begin{align}
  \label{eq:HKD}
  H_\gamma = & E^\gamma
          + \frac{1}{2} \mu_B \vec{B} \mathbf{g}_{\gamma} \vec{\sigma}_{\gamma}
          + \frac{1}{2} \vec{\sigma}_{\gamma} \mathbf{A}_\gamma \vec{I}
          + \mu_N g_N \vec{B}\cdot \vec{I}
\end{align}
where the index \(\gamma = (i,R)\) labels the KD that originates from the crystal-field orbital \(i=1,2,3\), transforms according to the representation \(R = \Gamma_4,\Gamma_{5/6}\),
and is made up by the pseudospin states \(\ket{\gamma,\sigma=\uparrow,\downarrow}\).
The Hamiltonian $H_\gamma$ for the KD $\gamma$ consists of its electronic zero-field energy \(E^{\gamma}\),
the Zeeman interaction coupling the pseudospin states of the KD to
the magnetic field \(\vec{B}\),
the hyperfine interaction coupling the pseudospin to the nuclear spin \(I\),
and the nuclear Zeeman term describing the coupling of the nuclear spin to the magnetic field.

The precise form of the hyperfine and \(g\)-tensors deviate from a simple spin model and depend on the KD~\cite{tissot21a,tissot21b}; their explicit form is given in Appendix~\ref{app:tensors} and summarized in the following.
We choose the \(z\)-axis \emph{parallel} to the stacking axis of the crystal
and
use the Pauli vector \(\vec{\sigma}_{\gamma}\) consisting of the standard Pauli operators \(\sigma_{\gamma}^k\) (\(k=x,y,z\)) acting between the pseudospin states \(\ket{\gamma,\sigma=\uparrow,\downarrow}\), and the nuclear spin operators \(I_k\) in units of the reduced Planck constant \(\hbar\).
The \(g\)-tensors are all diagonal, for \(\Gamma_{5/6}\) KDs only the \(z,z\)-component is allowed, and for the \(\Gamma_4\) KDs the \(x,y\) components have the same absolute value, with the same sign for the KD originating from the orbital singlet \(i=3\) and opposite signs for the doublet KDs \(i=j=1,2\).

We denote the parallel (perpendicular) \(g\)-factors of the KDs with \(g_{\gamma}^{z(x)}\). Perpendicular \(g\)-factors of the KDs originating from the orbital (\(j=1,2\)) doublets are, however, not considered since they either vanish due to symmetry (\(\Gamma_{5/6}\)) or are much smaller than the parallel component (\(j,\Gamma_4\), not experimentally resolved)~\cite{kaufmann97,gilardoni20,wolfowicz20,csore20,tissot21a}. 

The hyperfine coupling tensors for \(\Gamma_{5/6}\) only couple \(I_z\) to \(\sigma_{j,\Gamma_{5/6}}^{x}\) and \(\sigma_{j,\Gamma_{5/6}}^{z}\) with the coupling strength \(a_{\Gamma_{5/6}}^{x}\) and \(a_{\Gamma_{5/6}}^{z}\), respectively.
The coupling tensors for \(\Gamma_4\) KDs are diagonal and fulfil \(a_{j,\Gamma_4}^y = -a_{j,\Gamma_{4}}^x\) for the KDs from orbital doublets and \(a_{3,\Gamma_4}^y = a_{3,\Gamma_{4}}^x\) for the singlet KD~\cite{tissot21b},
where we denote the diagonal entries as \(a_{j,\Gamma_4}^k\) ($k=x,y,z$).
The different forms of the hyperfine coupling leads to different mixtures of nuclear and pseudospin levels inside the KDs.

Furthermore, we use the Bohr (nuclear) magneton \(\mu_B\) (\(\mu_N\)), and the nuclear \(g\)-factor \(g_N\).
Here we have \( | \mu_N g_N | \ll | \mu_B g_{i,\gamma}^{z} |\), in particular \(\mu_N g_N / \mu_B \approx 10^{-4}\) for V.


The electronic selection rules derived in \cite{tissot21a} are summarized and further refined to include the polarization of the perpendicular field in Fig.~\ref{fig:selection}(b);
enabling simple access to selection rules for circular polarization \(\vec{{F}}_{\pm} = F {(\cos \omega t, \pm \sin \omega t, 0)}\) with polarization \(\pm\), electric or magnetic field strength \(F=\mathcal{E},B\) and positive angular frequency \(\omega\).
The selection rules in Fig.~\ref{fig:selection}(b) correspond to non-zero matrix elements for \({F}_{\perp} = {F}_x + \mathrm{i} {F}_y\) which combined with the energy ordering leads to the selection rules for circular polarization, i.e. \(F_{\perp} \to F_- (F_+)\) for arrows in Fig.~\ref{fig:selection}(b) pointing from a lower to a higher (higher to lower) energy.
We stress that the level ordering can depend on the defect configuration
and note that the different forms of the hyperfine coupling tensors of the KDs [see \eqref{eq:HKD}]
are suited to assign the irreps to the physical states as was done in~\cite{tissot21b}. 
The depicted ordering in Fig.~\ref{fig:selection} corresponds to V$^{4+}$ \(\alpha\) defect in 4H-SiC.

As an example we consider an optical, resonant drive between the ground state (GS) \(g=(1,\Gamma_4)\)
and the exited state (ES) \(e=(2,\Gamma_{5/6})\) pseudospin manifold,
i.e. a drive with angular frequency \(\omega_d > 0\), leading to a transition matrix element of the form
\begin{align}
  \label{eq:polsel}
  \braket{ e, \uparrow | H_d | g, \uparrow } = e^{\mathrm{i} (E^{e,\uparrow} - E^{g,\uparrow}) t/\hbar} \epsilon \mathcal{E}_{\perp}(t)
  = \mathcal{E} \epsilon e^{\mathrm{i} \delta_{\pm} t},
\end{align}
with the dipole matrix element of the transition \(\epsilon\).
The detuning $\delta_{\pm} = (E^{e,\uparrow} - E^{g,\uparrow})/\hbar \pm \omega_d$ of the transition strongly depends on the polarization ``$\pm$'' of the drive.
Within the rotating wave approximation, only the ``-'' polarized drives with $\delta_- = 0$ remain.
Therefore, the selection rules in Fig.~\ref{fig:selection}(b) can be interpreted as circular polarization dependent selection rules, with the aforementioned mapping.
The total angular momentum for these atom-photon interactions is conserved because a change of pseudospin goes hand in hand with a change of electron spin as well as angular momentum (see \cite{tissot21a} for the form of the KD states).

In the following we will use a dipole moment of \(\epsilon = 1\,\)debye which was estimated in \cite{gilardoni21} based on the radiative lifetime of the defect for all leading order transitions and estimate the transition dipole elements \(\tilde{\epsilon} \sim \epsilon \Delta_{\mathrm{so}}^1/\Delta_{\mathrm{cr}} \approx 0.002 \epsilon\) for purely spin-orbit mixing allowed transitions.

To generalize the selection rules to include the nuclear spin, the simplest approach is to use the admixture of states,
given via the diagonalization of the static KD Hamiltonians.
Here we only present the levels relevant for the protocol introduced in the following.
For the GS we arrive at the unitary transformation
\begin{align}
  \label{eq:TG4}
  & T_{g} = \exp \left( - \hspace{-3.6mm} \sum_{m = -I+1}^{I} \hspace{-3.5mm} \theta_{m} \ket{g,\uparrow}\ket{m} \bra{g,\downarrow}\bra{m-1} - \mathrm{h.c.} \right),
\end{align}
with the nuclear spin state \(\ket{m} = \ket{I,m}\) and the mixing angles $\theta_{m}$ given by \(\tan(2\theta_{m})\! = \!\frac{a_{g}^{x} \sqrt{I(I+1) - m(m-1)}}{( \mu_B g_{g}^{z} + \mu_N g_N ) B + a_{g}^{z} (m - 1/2)}\).
For the ES \(\ket{e,\sigma} = \ket{2,\Gamma_{5/6},\sigma}\), used as an ancillary state manifold in the following, we find the transformation
\begin{align}
  \label{eq:TG56}
  T_{e} = \exp \left( - \sum_{m = -I}^{I} \phi_{m} \ket{e,\uparrow}\ket{m} \bra{e,\downarrow}\bra{m} - \mathrm{h.c.} \right),
\end{align}
with mixing angles $\phi_{m}$ given by \(\tan(2\phi_{m})~=~\frac{a_{e}^{x} m}{\mu_B g_{e}^{z} B + a_{e}^{z} m}\).
The corresponding energies are
\begin{align}
  \label{eq:EG4}
  E^{g, \sigma}_{m-\delta_{\sigma,\downarrow}} =
    & \sigma \frac{\mu_B g^{z}_{g} B + a^{z}_{g} (m - 1/2) + \mu_N g_N B}{2 \abs{\cos 2 \theta_{m}}} \notag \\
    & + \frac{a^{z}_{g}}{4} + \mu_N g_N B (m - 1/2), \\
  \label{eq:EG56}
  E^{e, \sigma}_{m} = \,
    & \Delta_{\mathrm{cr}}
    + \sigma \frac{\mu_B B g^{z}_{e} + a^{z}_{e} m}{2 \abs{\cos 2 \phi_{m}}}
      + \mu_N g_N m B,
\end{align}
where we choose  $E^g=0$ to lie at zero energy such that 
the energy of the ES corresponds to the crystal field splitting \(E^{e} = \Delta_{\mathrm{cr}}\) (the second GS and ES are offset by the spin-orbit splitting \(E^{1,\Gamma_{5/6}} = \Delta^1_{\mathrm{so}}\), \(E^{2,\Gamma_4} = \Delta_{\mathrm{cr}} + \Delta^2_{\mathrm{so}}\)).
We use the Kronecker symbol $\delta_{\sigma,\sigma'}=\left(1\text{ if }\sigma=\sigma'\text{ else }0\right)$ for compact notation.
From now on we label the eigenstate pertaining to \(E^{\gamma,\sigma}_m\) as \(\ket{\gamma,\sigma,m}\) according to the KD \(\gamma=g,e\), the main pseudospin component \(\sigma=\uparrow,\downarrow\), and the main nuclear magnetic quantum number \(m\).
In Fig.~\ref{fig:Egs} we plot the ground state spin multiplet energies of the Vanadium \(\alpha\) defect of 4H-SiC as a function of the parallel magnetic field strength \(B\).
For further details, we refer the reader to\ \cite{tissot21b} where the remaining KDs, higher orders of the hyperfine interaction, as well as the nuclear quadrupole interaction were considered as well.

\section{Qubit Design and State Preparation\label{sec:design}}
\subsection{Qubit Candidates\label{sec:qubits}}
\begin{figure}
\centering
\includegraphics[width=8.5cm]{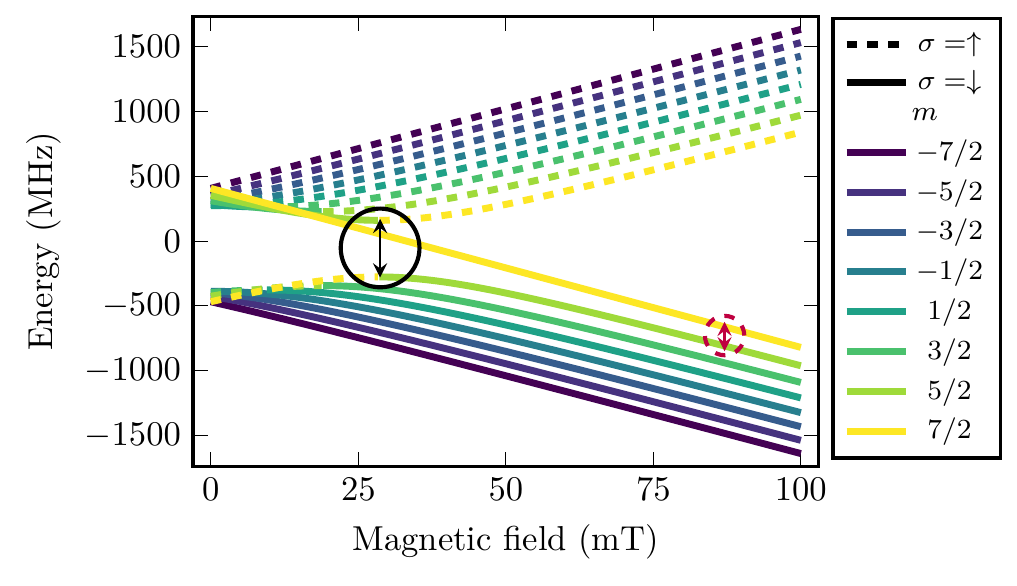}
\caption{\label{fig:Egs} Ground state [\(g=(1,\Gamma_4)\)] nuclear energy levels of the vanadium \(\alpha\) defect in 4H-SiC as a function of the magnetic field strength parallel to the crystal axis.
  The black solid and red dashed circles mark two possible zero first order Zeeman (ZEFOZ) transitions (see main text).
  These pairs of levels are suitable qubit candidates due to their enhanced protection from magnetic noise.
  The coupling to perpendicular fields is negligible.
  Solid (dashed) lines label states that consist mainly of pseudospin down (up) and the line color encodes the 
  nuclear spin quantum number $m$ (see legend).}
\end{figure}
Based on the outlined theory, we now discuss suitable qubit candidates.
Due to the time-reversal symmetry of the KDs, electric fields cannot lead to pseudospin flips and dephasing in the leading order.
For temperatures $\ll |\Delta_{\mathrm{so}}^1 / k_B| \approx 25\,$K for the considered $\alpha$ V ($10\,$K for Mo) 
transitions to the second GS ($1,\Gamma_{5/6}$) are suppressed, these would else limit the decoherence times~\cite{jahnke15,gilardoni20}.
Furthermore, recent experiments show that the $T_1$ exceeds seconds below $\lesssim 1\,$K \cite{astner22} ($4\,$K for Mo defects \cite{gilardoni20}),
such that we expect the main decoherence source to be coupling to fluctuations in the magnetic field due to the spin bath at low temperatures.
For this reason, we discuss two qubit candidates that are protected against magnetic noise in the following.

To achieve the protection we suggest the use of zero first order Zeeman (ZEFOZ) transitions, also known as clock transitions, or optimal working points or ``sweet spots" which are already established for different transition types and materials~\cite{fraval04,mohammady10,ma15,ortu18} including the divacancy defect in SiC~\cite{miao20,onizhuk21}.
Using the state pair belonging to such a transition as the qubit improves protection from magnetic or nuclear spin bath noise (by suppressing the first order of the coupling),
thereby offering the opportunity to increase the coherence time.
Due to the large anisotropy of the \(g\)-tensor, i.e.\ its vanishing perpendicular component as well as the orders of magnitude smaller coupling of the nuclear spin to the magnetic field in comparison to the coupling of the electronic state,
we optimize ZEFOZ transitions only via the parallel magnetic field component.
Inspecting the field dependence of the energies we find two conceptual possibilities of (approximate) ZEFOZ transitions with different strengths and weaknesses,
see Fig.~\ref{fig:Egs}.

The first possibility is an \emph{electronic spin qubit} (transition marked in black at  \(\approx 28\,\)mT in Fig.~\ref{fig:Egs}) at the point of an avoided crossing, using the levels \( \ket{g,\downarrow,5/2} \) and \( \ket{g,\uparrow,7/2} \).
At this point the Zeeman interaction  and the diagonal hyperfine coupling are of similar magnitude, leading to a high degree of mixing of the states.
The energy levels of the transition have an extremum as a function of magnetic field at this point, implying that they are parallel and constitute a ZEFOZ transition.
Additionally, their eigenstate characteristics enable strong microwave driving using a parallel microwave magnetic field, i.e.
\begin{align}
  \label{eq:MWD}
  \braket{ g,\uparrow,I | \sigma_{g}^z | g,\downarrow,I-1 } = - \sin(2 \theta_{I}),
\end{align}
which is \(-1\) at the avoided crossing.

The other possibility are \emph{nuclear spin qubits} given by neighbouring hyperfine levels within the same pseudospin manifold at higher magnetic fields.
For sufficiently high bias magnetic field, the field dependence of the transition frequency becomes negligible (see Fig.~\ref{fig:Egs}),
because the Zeeman splitting suppresses the (off-diagonal) hyperfine interaction, leaving only the small nuclear Zeeman term.
As is visible in Fig.~\ref{fig:Egs} there is a regime where the nuclear levels are approximately parallel (leading to ZEFOZ transitions) and addressable individually due to the large splitting between the hyperfine levels, as well as the sufficient anharmonicity.
This enables the possibility of higher-dimensional encoding in a single defect in this magnetic field domain.
Direct driving between the nuclear levels is only possible via small terms of the Zeeman Hamiltonian,
for example for the red transition in Fig.~\ref{fig:Egs} we have the leading element $\braket{g,\downarrow,I-1|H_z^\mathrm{MW}|g,\downarrow,I} \approx \sqrt{2I} \mu_N g_N B_{\perp}^{MW}(t)$ that can be driven using a perpendicular microwave magnetic field.

To compare the protection from magnetic noise we calculate the leading order energy fluctuation induced by a fluctuation of the magnetic field $\delta \vec{b}$~\cite{wolfowicz21}.
In both cases the immediate influence on the energies stems from $b_z$,
for the electronic ZEFOZ qubit we find $\delta E_{\mathrm{el}} \approx [ (\mu_B g_g^z + \mu_N g_N) \delta b_z]^2 / 2 \sqrt{2 I} a_g^x$, i.e. the first order truly vanishes and the second order is suppressed by the perpendicular hyperfine interaction.
For the nuclear (ZEFOZ like) qubit the leading order is given by $\mu_N g_N \delta b_z$, i.e. it does simply not include the much larger electronic term but it is still linear.
Both are large improvements over a na\"ive electronic qubit where the leading term would be $\mu_B g_g^z \delta b_z$.
We expect that these optimized transitions, at low temperature will enable a big improvement over the $T_2^* \approx 0.3\,\mu$s measured for Mo at $4\,$K in \cite{bosma18}.

The optical linewidth prevents pseudospin resolution at the avoided crossing as well as optical nuclear spin readout.
The combination of larger level splitting and Rabi frequencies of (non-ZEFOZ) electronic qubits,
could make a hybrid of a electronic qubit, for control and readout, and a nuclear qubit, for storage, an interesting option.
Finally, the hyperfine structure of all V defects in 4H and 6H-SiC suggests that all have the \((1,\Gamma_4)\) KD as the lowest GS, making the above arguments applicable to all defects in this family~\cite{wolfowicz20,tissot21b}.

\subsection{State Preparation via Nuclear Polarization\label{sec:pol}}
\begin{figure}
\centering
\includegraphics[width=8.5cm]{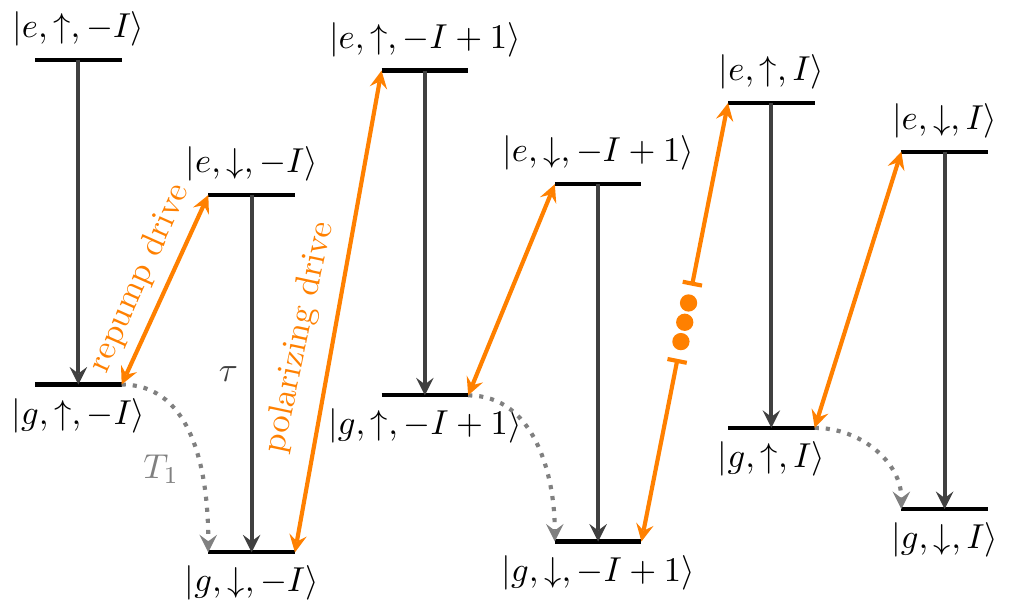}
\caption{\label{fig:scheme}
  Schematic of a purely optical (ratchet-type) nuclear polarization method.
  This scheme makes use of the short exited state lifetime \(\tau\).
  The different forms of the hyperfine coupling make it possible to use a set of pump fields to increase the nuclear polarization while exciting from the GS \(g=(1,\Gamma_4\)) pseudospin \(\downarrow\) manifold to the ES \(e=(2,\Gamma_{5/6}\)) \(\uparrow\).
  Another set of drives is used to repump the pseudospin \(\downarrow\) manifold while conserving the nuclear spin.
  Here both of the drive types are implemented using ``-'' polarization according to the selection rules depicted in the magnification in Fig.\ref{fig:selection}(b) in combination with the hyperfine mixing of the GS and ES.
  The repump drives are employed to avoid a bottleneck due to a small pseudospin relaxation  rate \(1/T_1\), and instead rely on the fast decay of the ES.
}
\end{figure}

For both the electron and nuclear spin qubits it is necessary to polarize the nuclear spin to achieve a well defined initial state mandatory for experiments and potential technological applications.
Due to the multitude of nuclear spin states for \(I=7/2\) in the case of V and \(I=5/2\) for Mo isotopes with nuclear spin,
we develop a dissipative nuclear-spin polarization protocol.
The dissipative nature makes it possible to use continuous drives,
rendering it unnecessary to measure and then manually choose the correct pulse
to make the process irreversible.

First, we outline the general idea and then  discuss one possible implementation in more detail.
The protocol relies on the different forms of the hyperfine coupling of the KDs~\eqref{eq:HKD}
to open different channels via the state mixing.
The first order nuclear spin \emph{polarizing}, pseudospin flipping transition,
is the leading one because the leading order (in spin-orbit coupling) allowed transitions are all pseudospin conserving, see Fig.~\ref{fig:selection}.
This makes a \emph{repump} necessary to repopulate the correct pseudospin manifold.
In summary, to polarize the nuclear spin inside a nuclear spin manifold of a KD, we employ an ancillary KD with a different form of the hyperfine coupling.
This ensures that we can engineer the driving such that the drive to and decay from the ancillary KD on average polarizes the nuclear spin.


Different ES, different transitions, or different decays can be employed.
For simplicity we concentrate on a purely optical protocol that
relies on a qubit transition in the GS KD \(g=(1,\Gamma_4)\) and using the ancillary ES manifold \(e=(2,\Gamma_{5/6})\) KD to prepare the final state \(\ket{g,\downarrow,I}\).
This corresponds to a crystal splitting of $\approx 234\,$THz (or a wavelength of $\approx 1.28\,\mu$m) for the V \(\alpha\) defect in 4H-SiC.
This ES has a short lifetime of \(\tau=167\,\)ns~\cite{wolfowicz20} in this configuration, making it particularly suited for a dissipative protocol.
We furthermore drive the polarizing transition due to better control of the drive compared to using a decay channel.
Lastly, we rely on the leading decay process, instead of additional channels due to the hyperfine mixing, thereby leading to faster dynamics.
Figure~\ref{fig:scheme} illustrates the main processes involved in the polarization  towards the final state \(\ket{g,\downarrow,I}\).

To avoid unnecessarily populating one of the other KDs the driving fields \(\mathcal{E}\) should fulfil \(|\epsilon \mathcal{E}| \ll \min_i|\Delta_{\mathrm{so}}^i|\), as is also required for the use of the rotating wave approximation.
A drive exceeding this limit would also exceed the breakdown electric field of SiC by orders of magnitude~\cite{yamaguchi18}.
A resonant optical drive can excite the system from the excited state to the conduction band, thereby ionizing the defect \cite{spindlberger19,wolfowicz20}.
The occupation of the excited states should therefore be minimal as well.
To this end, we conservatively limit the largest resonant Rabi-frequency to fulfil $\Omega_{R} \ll 1 /\tau \approx 6 \,(\mu$s$)^{-1}$, i.e. significantly below saturation.

Therefore we can restrict the driving Hamiltonian to the allowed transitions between GS and ES, see the magnified part in Fig.~\ref{fig:selection}(b),
leading to
\begin{align}
  \label{eq:Hd_pol}
  H_d^{\sigma} \approx \mathcal{E}(t) ( \epsilon \ket{e,-\sigma} + \tilde{\epsilon} \ket{e,\sigma})\bra{g,-\sigma}
  + \mathrm{h.c.}
\end{align}
in the product basis, where we now use \(\sigma=\pm=\uparrow,\downarrow\) to indicate the polarization as well as the pseudospin, enabling a compact encoding of the selection rules. 
Here, \(\mathcal{E}(t)\) is the time dependent electric field amplitude in the rotating wave approximation relevant for the \(g \to e\) transitions (oscillating with frequencies close to \(\Delta_{\mathrm{cr}}\)).

Combined with the hyperfine mixing of the states [see Eqs.~\eqref{eq:TG4}~and~\eqref{eq:TG56}],
where \(\ket{g,\downarrow}\ket{m}\) is mixed with \(\ket{g,\uparrow}\ket{m+1}\)
and \(\ket{e,\downarrow}\ket{m}\) is mixed with \(\ket{e,\uparrow}\ket{m}\),
this implies that using ``\(-\)'' polarization enables the \emph{polarizing} and \emph{repump} drives while suppressing most unwanted transitions.
For large Zeeman splittings, linear polarization can also be used, assuming that the individual transitions are spectrally resolved, because the larger detuning suffices to suppress unwanted transitions.

\begin{figure}
\centering
\includegraphics[width=8.5cm]{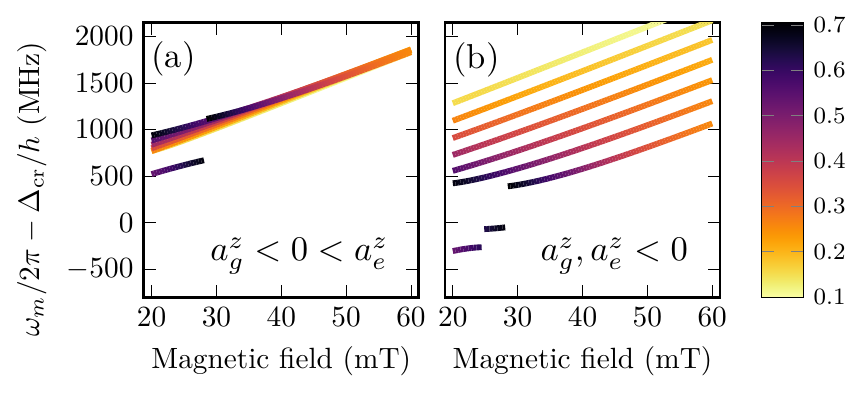}
\caption{\label{fig:transE}
  Polarizing transition frequencies \(\omega_{m}/2\pi = (E^{e,\uparrow}_{m+1} - E^{g,\downarrow}_{m})/h\) as a function of the magnetic field strength along the crystal axis.
  We show the significant difference between the cases (a) \(\mathrm{sign}(a_{g}^z) = \mathrm{sign}(a_{e}^z)\) and (b) \(\mathrm{sign}(a_{g}^z) = -\mathrm{sign}(a_{e}^z)\) for the driving frequencies.
  For drives corresponding to the correct frequencies, the protocol works in either case.
  The color shade gives the relative resonant driving strength \(\braket{ e,\uparrow,m+1 | H_d | g,\downarrow,m }/\epsilon \mathcal{E}(t)\).
  The jumps in the frequencies correspond to avoided crossings in the ES and GS manifolds where the states associated with the main nuclear and pseudospin components change.
}
\end{figure}
To achieve fast dynamics the driving field \(\mathcal{E}(t) = \sum_f \mathcal{E}_f e^{- \mathrm{i} \omega_f t}\) should consist of a resonant field for each of the transitions corresponding to circularly polarized drives with amplitudes \(\mathcal{E}_f\) and rotating with frequencies \(\omega_f\)).
The optimal angular frequencies of the drives are \(\omega_{m} = (E^{e,\uparrow}_{m+1} - E^{g,\downarrow}_{m})/\hbar\) with \(m=-I,\dots,I-1\) (\emph{polarizing}) and \(\omega_m' = (E^{e,\downarrow}_{m} - E^{g,\uparrow}_{m})/\hbar\) with \(m=-I,\dots,I\) (\emph{repump}) using the eigenenergies~\eqref{eq:EG4}~and~\eqref{eq:EG56}.
%
The magnetic field dependence of the polarizing frequencies \(\omega_m\) can be seen in Fig.~\ref{fig:transE},
including the relative transition matrix element  \(\braket{ e,\uparrow,m+1 | H_d | g,\downarrow,m }\!/\epsilon \mathcal{E}(t)\) given by the exact transformation~\eqref{eq:TG4}~and~\eqref{eq:TG56}.
The figure also shows the important role played by the signs of the hyperfine components \(a_{\gamma}^z\) ($\gamma=g,e$) which give the ordering of the nuclear states.
This ordering is currently unknown, but has a dramatic effect on the requirements of the polarizing scheme:
If \(a_{\gamma}^z\) have opposite signs in the two KDs, a single optical frequency can be sufficient to drive the pseudospin-flipping transition.

Assuming \(a_{g}^z a_{e}^z >0\), there are 15 different optical transitions (for V where $I=7/2$) that need to be driven in order to fully polarize the nuclear spin.
Such a complex excitation spectrum, containing 15 different laser frequencies, can be produced by direct, external, or combined modulation of a laser diode \cite{drever1983,schrenk2020,saliou2021}.
We note, however, that in this worst case scenario the spin polarization can also be achieved by driving the system with only two excitation frequencies at any one time: Once a pair of hyperfine states has been depleted, the driving frequencies can be shifted to the next step in the polarization ladder, since the leakage by decay across two nuclear states is expected to be negligible. For increased fidelity, an excitation with four drives can be employed.

Inhomogeneous broadening might appear to be deleterious for such a scheme since the hyperfine transitions cannot be addressed individually. While vanadium in SiC presents a very stable mean frequency, spectral diffusion leads to a single-transition linewidth of order $400\,$MHz \cite{wolfowicz20}. In fact, this setting somewhat simplifies the polarization task since, for sufficiently large bias fields, the repump and polarization transitions are spectrally separated from the spin conserving transition, while the remaining undesired transitions are suppressed by the choice of laser polarization (see Eq.~(\ref{eq:Hd_pol}) and Appendix~\ref{app:Hdeff}). 
 Given that the spectral span of the transitions is $\sim1.6\,$GHz, it will therefore be sufficient to apply four laser frequencies to address all desired transitions in one branch, or to sweep the repump and polarization laser frequencies at a rate slower than the expected nuclear spin transfer.

In this context, we highlight that the only possible unwanted transitions accessible with ``\(-\)'' polarization between the GS and ES are pseudospin conserving, i.e.\ coupling the pseudospin manifolds \(\uparrow \Leftrightarrow \uparrow\) (\(\propto \epsilon\)) and \(\downarrow \Leftrightarrow \downarrow\) (\(\propto \tilde{\epsilon} a_{g}^{x} \) and \(\propto \epsilon a_{g}^{x} a_{e}^{x}\)).
The combination of the spin-orbit mixing and both hyperfine mixings only leads to a correction of the amplitude of the polarizing transition from \(\ket{g, \downarrow, m} \Leftrightarrow \ket{e, \uparrow, m+1}\) (\(\propto \tilde{\epsilon} a_{g}^{x} a_{e}^{x}\)).
Most of these transitions do not drive away from the final state (discussed later) and therefore at most lead to a slow-down.
The problematic transition is  \(\ket{g,\downarrow,I} \Leftrightarrow \ket{e,\uparrow,I}\) which can interfere with the final state.
But as it is only allowed due to the combination of spin-orbit and hyperfine mixing (or the combination of the GS and ES hyperfine mixing), it is inefficient compared to the competing drive
\(\ket{g,\uparrow,I} \Leftrightarrow \ket{e,\downarrow,I}\) which is independently allowed due to the hyperfine mixing and spin-orbit mixing.



We model the dynamics imposed by the drive in combination with the decay of the ES using a Lindblad master equation
\begin{align}
  \label{eq:LBM}
  \dot{\rho} = \frac{i}{\hbar} \comm{\rho}{H} + \sum_{l} \Gamma_{l} \left( \sigma_l \rho \sigma_l^{\dag} - \frac{1}{2} \{ \sigma_l^{\dag} \sigma_l, \rho \} \right),
\end{align}
with the anticommutator \(\left\{ A, B \right\} = AB + BA\)
where we describe the optical decay with the dissipators \(\sigma^{\mathrm{op}}_{\uparrow/\downarrow} = \op{g,\uparrow/\downarrow}{e,\uparrow/\downarrow}\) both at a rate \(\Gamma_{\mathrm{tot}}=1/\tau\).
Moreover, it is possible to take  into account additional dissipation channels, such as the pseudospin relaxation \(\sigma_{\mathrm{rel}} = \op{g,\downarrow}{g,\uparrow}\) with rate \(1/T_1\) and decoherence \(\sigma_{\mathrm{ph}} = \op{g,\uparrow}{g,\uparrow}\) with rate \(1/T_2\).
The Hamiltonian \(H = H_0 + H_d\) consists of the static part \(H_0 = \bigoplus_{\gamma} H_{\gamma}\) made up by the KD Hamiltonians, as well as the driving Hamiltonian \(H_d\) [see Eq.~\eqref{eq:Hd_pol} for the relevant part].

Before discussing the precise dynamics we discuss a simplified model
using an effective Hamiltonian treating the hyperfine interaction as a perturbation using a first order Schrieffer-Wolff transformation~\cite{bravyi11}
as well as adiabatically eliminating the dynamics of the ES~\cite{reiter12}.
The details and derivation are given in Appendix~\ref{app:Hdeff}.
The leading order rates between states are
\begin{widetext}
  \begin{align}
  \label{eq:rates}
  \Gamma_{\uparrow,m \to \downarrow,m} = & \Gamma_{\mathrm{tot}} \left| \sum_f \mathcal{E}_f e^{- i \omega_f t} \left[\frac{\tilde{\epsilon} - {\epsilon a_e^x m}/{(2 g_e^z \mu_B B)}}{ {E^{e,\downarrow}_{m}}' - {E^{g,\uparrow}_{m}}' - \omega_f - \frac{\mathrm{i} \Gamma_{\mathrm{tot}} }{2} } -  \frac{\epsilon {a_e^x m}/{(2 g_e^z \mu_B B)}}{ {E^{e,\uparrow}_{m}}' - {E^{g,\uparrow}_{m}}' - \omega_f - \frac{\mathrm{i} \Gamma_{\mathrm{tot}} }{2} } \right] \right|^2 + \frac{1}{T_1} , \\
  \label{eq:ratesB}
  \Gamma_{\downarrow,m \to \uparrow,m+1} = & \Gamma_{\mathrm{tot}}  \left( \frac{\epsilon a_g^x }{2 g_g^z \mu_B B} \right)^2 [I(I+1)-m(m+1)]\left|\sum_f \frac{\mathcal{E}_f e^{- i \omega_f t}}{{E^{e,\uparrow}_{m+1}}' - {E^{g,\downarrow}_{m}}' - \omega_f - \frac{i \Gamma_{\mathrm{tot}} }{2}}\right|^2 ,
\end{align}
and the rate that needs to be suppressed for efficient nuclear-spin polarization is
\begin{align}
  \label{eq:badRate}
  \Gamma_{\uparrow,m \to \downarrow,m-1} = & \Gamma_{\mathrm{tot}}  \left( \frac{\epsilon a_g^x}{2 g_g^z \mu_B B} \right)^2 [I(I+1)-m(m-1)] \left|\sum_f \frac{\mathcal{E}_f e^{- i \omega_f t}}{{E^{e,\uparrow}_{m}}' - {E^{g,\uparrow}_{m}}' - \omega_f - \frac{i \Gamma_{\mathrm{tot}} }{2}}\right|^2 ,
\end{align}
\end{widetext}
where the \({E^{\gamma,\sigma}_m}'\) are the eigenenergies up to second order in the hyperfine coupling.
These rates encode the second order processes given by the driving to the ES followed by a decay to the GS as well as pseudospin relaxation with rate \(1/T_1\) relevant for weak repump drives.
We stress that tuning the driving frequencies \( \omega_f \) to the desired spin flip transitions minimizes the detuning in the denominator of the first two rates Eqs.~(\ref{eq:rates}) and (\ref{eq:ratesB}),
while leading to a denominator of the order of the Zeeman splitting in Eq.~\eqref{eq:badRate}, thus suppressing the last rate.

With this simplified model we can already understand a single cycle of the process (see Fig.~\ref{fig:scheme}) in terms of the second order processes leading to the effective rates.
An arbitrary nuclear state of the pseudospin down GS multiplet \(\ket{g,\downarrow,m \neq I} \) is driven to the pseudospin flipped and nuclear spin increased ES \(\ket{e,\uparrow,m+1}\) (see Fig.~\ref{fig:transE} for the frequencies of this transition as a function of the magnetic field)
and subsequently decays to \(\ket{g,\uparrow,m+1}\).
Because the lifetime \(T_1\) of the pseudospin is much longer than \(\tau \ll T_1\) the repump drive is used to transfer \(\ket{g,\uparrow,m+1}\) back to the \(\downarrow\) manifold conserving the nuclear spin via the pseudospin flipped state \(\ket{e,\downarrow,m}\) of the ES.
In the most likely case after this the state is \(\ket{g,\downarrow,m+1}\).
Repeating the cycles (i.e.\ letting the system evolve long enough) therefore drives the overall state to the final state \(\ket{g,\downarrow,I}\).

\begin{figure}
\centering
\includegraphics[width=8.5cm]{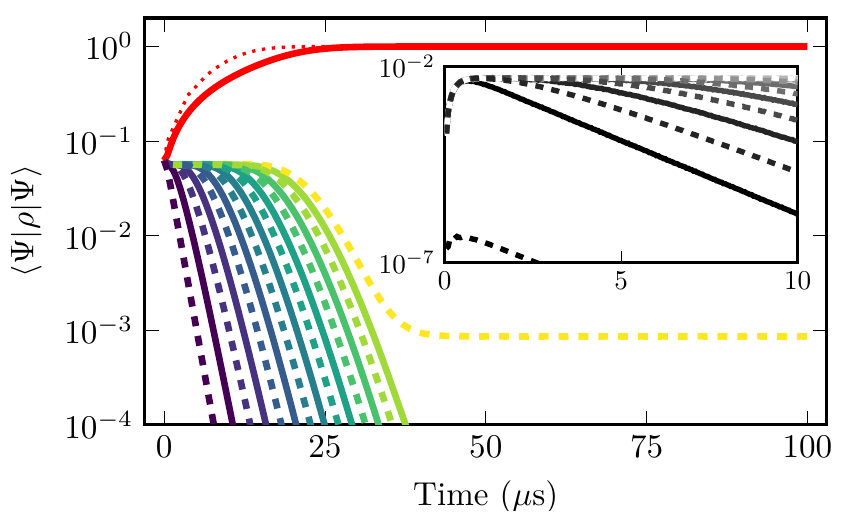}
\caption{\label{fig:dynamics}
  Time evolution of the level occupation probabilities during polarization pumping. 
  The states \(\ket{\Psi} = \ket{e/g,\sigma,m}\) are encoded in the line style,
  where blue and green (grey) shades indicate the main nuclear spin component of GS (ES), the dashed (solid) lines correspond to mainly pseudospin up (down) states, see legend of Fig.~\ref{fig:Egs} for the GS.
  For the ES, darker shades correspond to smaller \(m\), and the intermediate state is not visibly occupied. 
  The red line  corresponds to the desired final state \(\ket{g,\downarrow,I}\). The dotted red line corresponds to the analytic approximation (see Appendix~\ref{app:Hdeff}) to the simplified rate model [Eqs.~\eqref{eq:rates} and \eqref{eq:ratesB}].
  We only take the resonant parts of the rates into account,
  leading to the effective repump and polarizing rate $\Gamma_{\mathrm{eff}} = 4 \Omega_r^2 / \Gamma_{\mathrm{tot}}$.
  For the simulation we chose the drives such that all wanted transitions (see Fig.~\ref{fig:scheme}) have a resonant Rabi frequency $\Omega_r = 2\pi \cdot 0.2\,$MHz.
  These Rabi frequencies ensure that the ES is not significantly populated (see inset) because they are much smaller than the ES relaxation rate $\Gamma_{\mathrm{tot}} = 1 / 167\,$ns.
  See Appendix~\ref{app:param} for a discussion of all used parameters.
}
\end{figure}
In addition to the simplified model, we implemented the proposed protocol in the Julia programming language and numerically solved the dynamics of the density matrix~\cite{rackauckas17}.
For the numerical implementation we did not only use the leading order of the driving but kept all terms oscillating slower than \(2\,\)GHz.

Additionally, we included the intermediate KD \(\ket{1,\Gamma_{5/6},\sigma}\), and the lifetime, thermal excitation, as well as decoherence of the pseudospin.
If the lifetime of the intermediate state (IS) is much larger than \(\tau\) it can slow down the process.
This can be avoided with a repump excitation that drives back to the ES down state (\(\ket{1,\Gamma_{5/6},\sigma} \to \ket{e,\downarrow}\)).
On the other hand if either the decay rate to the IS is negligible or the lifetime of the IS is shorter than of the ES \(\tau\), it can speed up the polarization process by introducing an additional nuclear spin conserving decay channel.

Using the Kubo-Martin-Schwinger condition for the coupling to a heat bath~\cite{breuer02}, we can use the detailed balance to obtain \(\Gamma_{g,\downarrow \to g,\uparrow} = e^{- g_{g}^{z} \mu_B B / k_B T} / T_1 \approx 0.3 / T_1\) for \(T=100\,\)mK and $B=100\,$mT and using the Boltzmann constant \(k_B\).
These cryogenic temperatures are used in current experimental setups and are most likely necessary in potential applications for quantum technology.
We did not take into account thermal excitations between KDs,
because the large spin-orbit and crystal splitting protect against thermal excitations to higher energy KDs.
We expect that the pseudospin lifetime
(already measured for Mo~\cite{gilardoni20} and also expected for V \cite{astner22}) is much larger at low temperatures compared to the one used in the simulation.
Therefore, the lifetime becomes even less relevant compared to the fast dynamics due to the short ES lifetime,
see inset of Fig.~\ref{fig:dynamics} for the timescale.

We plot the resulting state occupation probabilities of the eigenstates of the static Hamiltonian at \(100\,\)mT as a function of time in Fig.~\ref{fig:dynamics} (see Appendix~\ref{app:param} for the remaining parameters). The additional effects we consider in the simulation are discussed in the following.

After \(100\,\mu\)s the fidelity of the final state \(\braket{g,\downarrow,I | \rho(100\,\mu\text{s}) | g,\downarrow,I} > 0.999 \).
We also note that a control field with different amplitudes of the different fields can achieve faster dynamics while still ensuring charge stability of the defect, because the matrix elements differ for each of the resonant transitions.
Finally, the inset of Fig.~\ref{fig:dynamics} shows that the ES does not have significant occupation.

Numerical simulations show that although the approximations necessary to derive the rates Eqs.~\eqref{eq:rates}--\eqref{eq:badRate} break down close to the avoided crossing (in our case at $28\,\mathrm{mT}$, see Fig.~\ref{fig:Egs}), the polarization protocol still works.
To use the electronic ZEFOZ transition in this case, NMR control can be used to transfer the population of the final state $\ket{g,\downarrow,I}$ to one of the levels involved in the qubit, i.e $\ket{g,\uparrow,I},\ket{g,\downarrow,I-1}$.

\subsubsection{Protocol Engineering\label{sec:engineering}}

While the outlined all optical protocol relies on the availability of circular polarization as well as an optical linewidth sufficiently narrow to resolve pseudospin transitions,
similar protocols using microwaves can be favorable depending on the technical constraints.
For example, if the optical linewidth is too broad to resolve the pseudospin transitions, using ``\(-\)'' polarization to mainly drive the pseudospin conserving transition \(g,\uparrow \Leftrightarrow e,\uparrow\) (for adequate bias magnetic field) is still possible.
In combination with a microwave with parallel polarization between the states \(\ket{g,\downarrow,m} \Leftrightarrow \ket{g,\uparrow,m+1}\) for \(m=-I,\dots,I-1\) analogous to Eq.~\eqref{eq:MWD}, the hyperfine allowed decay \(\ket{e,\uparrow,m+1} \Rightarrow \ket{g,\downarrow,m+1}\) then leads to the same final state \(\ket{g,\downarrow,I}\).
We expect this process to be slower than the outlined protocol as it relies on a decay process that is only allowed due to the hyperfine interaction, and has competing rates from the ES.

For a protocol that is less reliant upon circular polarization,
the same microwave drives are employed but now combined with an optical drive that is pseudospin flipping between \(\ket{g,\uparrow,m} \Leftrightarrow \ket{e,\downarrow,m}\) (the repump drive of the optical protocol),
like the all optical protocol this requires an optical linewidth that resolves the pseudospins.
This protocol also results in the same final state \(\ket{g,\downarrow,I}\).

The different available protocols highlight the prospect of engineering the protocols for these defects;
setting this work apart from studies that only investigated resolved polarization schemes, such as for erbium \cite{ranvcic17,stuart21}.

\subsubsection{Polarization Measurement\label{sec:measurement}}

We now briefly outline a measurement protocol to confirm the nuclear polarization.
The strongest optical transitions are pseudospin conserving ones and we expect the same to hold for the corresponding ES to GS decay at sufficiently high magnetic fields,
 as was confirmed in recent experiments~\cite{astner22}.
We therefore expect that the pseudospin state can be readout using a cycling transition, used in many platforms for readout~\cite{robledo11,delteil14,sukachev17,raha20,appel21} between the ES and GS, for example to readout the $\downarrow$ pseudospin a ``$+$'' polarized drive would be used.
However, even at moderate magnetic fields, the frequencies of these transitions are closely spaced or even overlapping for different hyperfine ground states. The spectral separation between electron spin-conserving transitions increases monotonically for bias field strengths greater than $\sim 40\,$mT, by approximately $6\,$GHz/T. Reading out the electron spin in this regime would then enable the detection of hyperpolarization by performing a hyperfine-selective electron spin $\pi-$rotation, i.e. a CNOT-like gate: A system that is initialized in one electronic state, but mixed across all nuclear spin states therein, would at best present an average contrast of $12.5\,\%$ for V, while the contrast for a perfectly initialized system would increase eight-fold and approach unity.

\section{Conclusions\label{sec:conclusion}}

Based on the current knowledge about TM defects in SiC we
found promising qubit candidates in the GS manifold, and developed the theory to engineer and quantitatively describe state preparation protocols.
We found that using the decay of an excited state in combination with drives enables dissipative polarization of the nuclear spin.
The main ingredients of our proposed protocol are a \emph{polarizing} transition, either a drive or a decay, enabled due to the different forms of the hyperfine coupling of the KDs
and a \emph{repump} drive that is used to induce pseudospin flips during the polarization process.
We applied this to the particular configuration of \(\alpha\) V defects in SiC using a purely optical protocol for the polarization.
Due to the plethora of nuclear states of these defects, nuclear spin polarization is essential for experimental implementations and quantum technology applications.

Considering the different options for the in-detail implementation of the protocol and
the fact that it is not necessary to address all transitions at the same time but gradually sweeping two drives is possible instead,
we estimate that the polarization can be achieved in state of the art experiments.
The sweep over drive frequencies can also be used to polarize a sub-ensemble of multiple similar defects.
%
For future research, it would be interesting to study the difference of the initialization of single defects and ensembles as well as measure additional rates for the different processes in these defects.
We deem the domain of static magnetic field \(\sim 100\,\)mT to resolve the pseudospin transitions favorable.
Experiments at temperatures \(\lesssim 1\,\)K where detrimental transitions to intermediate states and ES are suppressed and the $T_1$ is enhanced ought to benefit most from the ZEFOZ transitions.


\begin{acknowledgments}
  We thank C. Gilardoni for inspiring discussions and
  acknowledge funding from the European Union’s Horizon 2020 research and innovation programme under grant agreement No 862721 (QuanTELCO).
\end{acknowledgments}

\appendix
\section{\texorpdfstring{$g$}{g} and Hyperfine Tensors\label{app:tensors}}
In this appendix we give the explicit form of the coupling tensors used in the KD Hamiltonians~\eqref{eq:HKD} of the main text
\begin{widetext}
\begin{align}
  \mathbf{g}_{j,\Gamma_4} = \begin{pmatrix} g_{j,\Gamma_{4}}^{x} & 0 & 0 \\ 0 & -g_{j,\Gamma_{4}}^{x} & 0 \\ 0 & 0 & g_{j,\Gamma_{4}}^{z} \end{pmatrix}, \quad
  \mathbf{g}_{j,\Gamma_{5/6}} = \begin{pmatrix} 0 & 0 & 0 \\ 0 & 0 & 0 \\ 0 & 0 & g_{j,\Gamma_{5/6}}^{z} \end{pmatrix}, \quad
  \mathbf{g}_{3,\Gamma_4} = \begin{pmatrix} g_{3,\Gamma_{4}}^{x} & 0 & 0 \\ 0 & g_{3,\Gamma_{4}}^{x} & 0 \\ 0 & 0 & g_{3,\Gamma_{4}}^{z} \end{pmatrix}, \\
  \mathbf{A}_{j,\Gamma_4} = \begin{pmatrix} a_{j,\Gamma_{4}}^{x} & 0 & 0 \\ 0 & -a_{j,\Gamma_{4}}^{x} & 0 \\ 0 & 0 & a_{j,\Gamma_{4}}^{z} \end{pmatrix}, \quad
  \mathbf{A}_{j,\Gamma_{5/6}} = \begin{pmatrix} 0 & 0 & a_{j,\Gamma_{5/6}}^{x} \\ 0 & 0 & 0 \\ 0 & 0 & a_{j,\Gamma_{5/6}}^{z} \end{pmatrix}, \quad
  \mathbf{A}_{3,\Gamma_4} = \begin{pmatrix} a_{3,\Gamma_{4}}^{x} & 0 & 0 \\ 0 & a_{3,\Gamma_{4}}^{x} & 0 \\ 0 & 0 & a_{3,\Gamma_{4}}^{z} \end{pmatrix}.
\end{align}
\end{widetext}
Further details on the derivation can be found in \cite{tissot21a,tissot21b}.
\section{Model Parameters\label{app:param}}
\begin{table}
  \caption{\label{tab:par}
    Model parameters for the relevant KDs for the Vanadium \(\alpha\) defect in SiC.
    These parameters are based on fits \cite{tissot21b} to experimental data \cite{wolfowicz20} for the GS and experimental results \cite{astner22} for the ES.
    We use \(a_{e}^z < 0\) everywhere apart from Fig.~\ref{fig:transE}(a) where we use the opposite sign.
  }
  \begin{ruledtabular}
    \begin{tabular}{l r r r r}
          KD \(\gamma\) & \(E^{\gamma} / h\) (GHz) & \(g_{\gamma}^z\) & \(a_{\gamma}^z/h\) (MHz) & \(a_{\gamma}^{x}/h\) (MHz) \\
\hline
   \(1,\Gamma_{4}\) (\(g\)) & 0 & 1.748 & -232 & 165 \\
   \(1,\Gamma_{5/6}\) & 529 (\(\Delta_{\mathrm{so}}^1\)) & 2.16 & 170 & 210 \\
   \(2,\Gamma_{5/6}\) (\(e\)) & 234432 (\(\Delta_{cr}\)) & 2.18 & \(\mp\)213 & 75 \\
    \end{tabular}
  \end{ruledtabular}
\end{table}
In this article we use parameters according to estimates suited to describe the Vanadium \(\alpha\) defect in 4H-SiC. Other defects with the same electronic configuration can be treated analogously but the parameters will vary.
These parameter values are based on fits \cite{tissot21b} to experimental data \cite{wolfowicz20} for the GS as well as experimental data for the ES and $T_1$ time \cite{astner22}.
The parameters of the individual KDs can be found in Table~\ref{tab:par}
and the remaining relevant parameters are
the spin-orbit splitting of the ES KDs \(\Delta_{\mathrm{so}}^2/h=181\,\)GHz,
the nuclear gyromagnetic factor \(\mu_N g_N / h = -11.213\,\)MHz/T of Vanadium,
as well as the Bohr magneton \(\mu_B\).

To model the dissipative processes we use the measured lifetime (inverted rates) of the ES (\(2,\Gamma_{5/6}\)) \(\tau = 1/\Gamma_{\mathrm{tot}} = 167\,\)ns~\cite{wolfowicz20}.
Additionally we use a conservative estimate a spin lifetime \(T_1 = 500\,\mu\)s, and coherence time \(T_2 = 1\,\mu\)s,
that the spin-flipping decay from the ES to the GS at a rate \(|\Delta_{so}^1/(\Delta_{cr} \tau)|\), and to show that a decay over the \(\ket{1,\Gamma_{5/6},\sigma}\) does not interfere with the process the rates \(\Gamma_{e \to 1,\Gamma_{5/6}} = 2\,(\mu\)s$)^{-1}$ and \(\Gamma_{1,\Gamma_{5/6} \to g} = 10\,(\mu\)s$)^{-1}$.
As stated in the main text we also consider the inverted rate
\(\Gamma_{g,\downarrow \to g,\uparrow} = e^{- g_{g}^{z} \mu_B B / k_B T} / T_1 \approx 0.3 / T_1\) at \(T=100\,\)mK, with $B = 100\,$mT, and using the Boltzmann constant \(k_B\).

For the drive we used \(\epsilon = 1\,\)debye \cite{gilardoni21} for all leading order transitions and estimate the transition dipole elements \(\tilde{\epsilon} \sim \epsilon \Delta_{\mathrm{so}}^1/\Delta_{\mathrm{cr}} \approx 0.002 \epsilon\) for purely spin-orbit mixing allowed transitions.
The electric field amplitudes are chosen such that all resonant Rabi frequencies are $2\pi \cdot 0.2\,$MHz;
this corresponds to field strength $\mathcal{E}$ between $0.28\,$V$/$mm and $7.16\,$V$/$mm.
This ensures that the differences in the dipole elements (see Fig.~\ref{fig:transE}) do not lead to unnecessary bottlenecks in the protocol.
%
In the simulation we neglect transition matrix elements with a frequency above the cutoff frequency \(2\,\)GHz in the rotating frame.

\section{Derivation of the Effective Driving Hamiltonian\label{app:Hdeff}}
The selection rules (see Fig.~\ref{fig:selection}) imply that the transitions between the GS (1,\(\Gamma_4\)) and ES (2,\(\Gamma_{5/6}\)) KDs can be driven with perpendicular polarization.
Assuming a drive tuned to this transition we can neglect the off-resonant terms that would drive between other KDs
if the dipole element fulfills \(\abs{\epsilon \mathcal{E}(t)/\min_i |\Delta_{\mathrm{so}}^i|}\ll 1\).
Furthermore, applying a rotating wave approximation to neglect terms that oscillate with a frequency of about twice the transition frequency,
yields the circular polarization dependent driving Hamiltonian~\eqref{eq:Hd_pol}
\(H_d^{\sigma} \approx \mathcal{E}(t) ( \epsilon \ket{e,-\sigma} + \tilde{\epsilon} \ket{e,\sigma})\bra{g,-\sigma} + \mathrm{h.c.}\)
where \(\sigma=\pm=\uparrow,\downarrow\) indicates the polarization as well as the pseudospin, thereby encoding the selection rules,
\(\mathcal{E}(t) = \sum_{f} \mathcal{E}_{f} e^{- \mathrm{i} \omega_f t}\) is the time-dependent electric field amplitude in the rotating wave approximation for transitions from the GS to the ES resulting from circular drives with amplitudes \(\mathcal{E}_f\) and frequencies \(\omega_f\), and \(\epsilon\) (\(\tilde{\epsilon}\)) the leading order (spin-orbit mixing allowed) dipole matrix element of the transition.
For simplicity instead of the analytic diagonalization~\eqref{eq:TG4}~and~\eqref{eq:TG56}
we treat the hyperfine interaction as a perturbation compared to a large Zeeman splitting, i.e. \(B \gg \max_{\gamma} | \frac{a_{\gamma}^x}{\mu_B g_{\gamma}^z} | \approx 6.7\,\)mT.
For a more compact notation we use the ladder operators \(S_{\pm} = S_x \pm i S_y\) with \(S \in I,\sigma\) in the following.
And then diagonalize the static Hamiltonian using a first order Schrieffer-Wolff transformation~\cite{bravyi11}
\begin{align}
  \label{eq:SWT}
  & S_{g} 
           = \frac{a_{g}^{x} ( \sigma_{g}^+ I_+ - \sigma_{g}^- I_- )}{4 \mu_B g_{g}^{z} B},
  & S_{e} = \frac{ \mathrm{i} a_{e}^{x} \sigma_{e}^y I_z }{ 2 g_{e}^{z} \mu_B B } ,
\end{align}
leading to the transformed static Hamiltonian (energies up to second in \(a_{\gamma}^{x}\))
\begin{align}
  H_{g}' = & H_{g}\vert_{a_g^{x}=0} + \frac{1}{2} \comm{S_{g}}{\frac{a^{x}_g}{4} ( \sigma_g^+ I_+ + \sigma_g^- I_- )} \notag \\
  = & E^{g} + g_{g}^{z} \mu_B B \sigma_{g}^z / 2 + a_{g}^{z} \sigma_{g}^z I_z / 2 + \mu_N g_N I_z B  \notag \\
  \label{eq:HG4eff}
& + [ {a_{g}^{x}}^2/(4 g_{g}^{z} \mu_B B) ] \sigma_{g}^z [I(I+1) - I_z^2] \\
  H_{e}' = & H_{e}\vert_{a_e^{x}=0} + \frac{1}{2} \comm{S_{e}}{\frac{a^{x}_e}{2} \sigma_e^x I_z } \notag \\
= & E^{e} + g_{e}^{z} \mu_B B \sigma_{e}^z / 2 + a_{e}^{z} \sigma_{e}^z I_z / 2 + \mu_N g_N I_z B \notag \\
  \label{eq:HG56eff}
  & + \frac{{a_{e}^{x}}^2}{4 g_{e}^{z} \mu_B B} \sigma_{e}^z I_z^2
\end{align}
and the transformed driving Hamiltonian for a single polarization \(\sigma\)
\begin{widetext}
\begin{align}
  \label{eq:Hddash}
  H_d^{\sigma'} \approx &
  \mathcal{E}(t) ( ( \epsilon \ket{e,-\sigma} + \tilde{\epsilon} \ket{e,\sigma})\bra{g,-\sigma} + \comm{S_g + S_e}{( \epsilon \ket{e,-\sigma} + \tilde{\epsilon} \ket{e,\sigma})\bra{g,-\sigma}} ) + \mathrm{h.c.} \\
  = & \mathcal{E}(t) \big[ (\epsilon + \frac{-\sigma \tilde{\epsilon} a_{e}^x}{2 g_{e}^{z} \mu_B B} I_z) \ket{e,-\sigma}\bra{g,-\sigma} + (\tilde{\epsilon} + \frac{\sigma \epsilon a_{e}^x}{2 g_{e}^{z} \mu_B B} I_z ) \ket{e,\sigma}\bra{g,-\sigma}  \notag \\
  & + \frac{\sigma a_{g}^{x}}{2 \mu_B g_{g}^{z} B} I_{-\sigma} ( \epsilon \ket{e,-\sigma} + \tilde{\epsilon} \ket{e,\sigma} ) \bra{g,\sigma}  \big]
    + \mathrm{h.c.} .
\end{align}
In the following, we consider driving only with \(\sigma=-\) polarization, two sets of frequencies (one for the polarization and one for repumping),
and neglecting all decays apart from the very fast relaxation from the ES to the GS.
We use the theory from~\cite{reiter12} to eliminate the ES dynamics and derive effective dynamics of the GS.
For each of the frequencies we have
\begin{align}
  \label{eq:Hdf}
  H_d^f e^{i \omega_f t} / \mathcal{E}_f = & \epsilon \ket{e,+}\bra{g,+} + (\tilde{\epsilon} - \frac{\epsilon a_{e}^x}{2 g_{e}^{z} \mu_B B} I_z ) \ket{e,-}\bra{g,+} - \frac{a_{g}^{x}}{2 \mu_B g_{g}^{z} B} I_{+} \epsilon \ket{e,+}\bra{g,-}
\end{align}
the part of the driving Hamiltonian that drives from the GS to the ES.
The parts can be combined to the total driving Hamiltonian \(H_d = \sum_f (H_d^f + \mathrm{h.c.})\).
In combination with the dissipator \(L = \sqrt{\Gamma_{\mathrm{tot}}} \sum_{\sigma} \ket{g,\sigma} \bra{e,\sigma}\) which is transformed with the Schrieffer-Wolff transformation to
\(L' = \sqrt{\Gamma_{\mathrm{tot}}} \sum_{\sigma} \left( \ket{g,\sigma} - \sigma (\frac{a_g^x}{2 \mu_B g_g^z B} I_{-\sigma} + \frac{a_e^x}{2 g_e^z \mu_B B} I_z ) \ket{g,-\sigma} \right) \bra{e,\sigma} \),
we find the non-hermitian Hamiltonian \({H}_{\mathrm{NH}} = H_e' - \frac{\mathrm{i}}{2} \hbar {L}^{\dag} {L}\) where we neglected terms proportional to \((\mu_B g_{\gamma}^z B)^{-2}\).
Using this diagonal matrix we can apply~\cite{reiter12} and obtain
\begin{align}
{H}_{\mathrm{eff}} = & H_g' - \frac{1}{2} \left[ \sum_{f'} H_d^{f'}(t)^{+} \sum_{f,\sigma,m} \left( H_{\mathrm{NH}} - {E^{g,\sigma}_{m}}' - \hbar \omega_f \right)^{-1} H_d^{f}(t) \proj{g,\sigma,m} + \mathrm{h.c.} \right] \\
{L}_{\mathrm{eff}} = & L' \sum_{f,(\sigma,m)} \left( H_{\mathrm{NH}} - {E^{g,\sigma}_{m}}' - \hbar \omega_f \right)^{-1} H_d^{f}(t) \proj{g,\sigma,m}
\end{align}
where \({E^{\gamma,\sigma}_m}'\) are the diagonal entries of \(H_{\gamma}'\).

In accordance with the former Schrieffer-Wolff transformation, we neglect terms that are quadratic suppressed by the Zeeman splitting
(either directly or in terms of a rotating wave approximation).
We additionally treat \(\tilde{\epsilon}\) terms in the same way
and neglect off-diagonal (w.r.t. pseudospin of the KD) of this order, because they lead to higher order contributions.
This yields the simplified effective Hamiltonian
\begin{align}
  \label{eq:Heff}
H_{\mathrm{eff}} &
  \approx H_g' - \sum_{m} \frac{1}{2} \sum_{f',f} \Bigg[
    \frac{\mathcal{E}_{f'} \mathcal{E}_f e^{i (\omega_{f'} - \omega_f) t} \epsilon^2}{ {E^{e,\uparrow}_{m}}' - {E^{g,\uparrow}_{m}}' - \hbar \omega_f - \frac{\mathrm{i} \hbar \Gamma_{\mathrm{tot}} }{2} } + \mathrm{h.c.} \Bigg] \op{g,+,m}{g,+,m}
\end{align}
and the effective Lindblad operator
\begin{align}
   L_{\mathrm{eff}}' = &
\sqrt{\Gamma_{\mathrm{tot}} } \sum_{f,m} \mathcal{E}_f e^{- \mathrm{i} \omega_f t} \Bigg\{
\frac{\epsilon}{ {E^{e,\uparrow}_{m}}' - {E^{g,\uparrow}_{m}}' - \hbar \omega_f - \frac{\mathrm{i} \hbar \Gamma_{\mathrm{tot}} }{2} } \proj{g,+,m} \notag \\
& - \frac{a_g^x \sqrt{I(I+1)-m(m-1)}}{2 \mu_B g_g^z B} \frac{\epsilon}{ {E^{e,\uparrow}_{m}}' - {E^{g,\uparrow}_{m}}' - \hbar \omega_f - \frac{\mathrm{i} \hbar \Gamma_{\mathrm{tot}} }{2} }  \op{g,-,m-1}{g,+,m} \notag \\
& + \left[\frac{\tilde{\epsilon} - {\epsilon a_e^x m}/{(2 g_e^z \mu_B B)}}{ {E^{e,\downarrow}_{m}}' - {E^{g,\uparrow}_{m}}' - \hbar \omega_f - \frac{\mathrm{i} \hbar \Gamma_{\mathrm{tot}} }{2} } - \frac{\epsilon {a_e^x m}/{(2 g_e^z \mu_B B)}}{ {E^{e,\uparrow}_{m}}' - {E^{g,\uparrow}_{m}}' - \hbar \omega_f - \frac{\mathrm{i} \hbar \Gamma_{\mathrm{tot}} }{2} } \right] \op{g,-,m}{g,+,m} \notag \\
  \label{eq:Leff}
& - \frac{ \frac{\epsilon a_g^x}{2 \mu_B g_g^z B} \sqrt{I (I+1) - m (m+1)}}{ {E^{e,\uparrow}_{m+1}}' - {E^{g,\downarrow}_{m}}' - \hbar \omega_f - \frac{\mathrm{i} \hbar \Gamma_{\mathrm{tot}} }{2} } \op{g,+,m+1}{g,-,m} \Bigg\} .
\end{align}
The first term leads to a decoherence of the ES that is irrelevant for our protocol.
Because our protocol drives the spin flipping transitions, and the depolarizing term in the second line stems from the (off-resonant) spin-conserving excitation, it is naturally suppressed in comparison to the terms in the last two terms that lead to the polarization.
This suppression is effective because the detuning \({E^{e,\uparrow}_m}' - {E^{g,\uparrow}_m}' - \hbar \omega_f\) is of the same order as the Zeeman splitting,
and therefore this term is much smaller than the last two terms, where the detuning is small (or zero).

This means in the leading order we have the time-dependent transition rates shown in Eqs.~\eqref{eq:rates}--\eqref{eq:badRate}.
Here we see that in the ideal case every term of the sum in the rate~\eqref{eq:badRate} is suppressed by the Zeeman splitting to the power of four.
We furthermore re-included the inverse lifetime to eq.~\eqref{eq:rates} to highlight that the finite lifetime of the states does not work against our protocol but can enhance its performance for weak repump drives.

For appropriate drives we can neglect the unwanted terms in Eq.~\eqref{eq:Leff} and use that the effective Hamiltonian~\eqref{eq:Heff} is diagonal.
If we additionally neglect all terms oscillating with a frequency bigger than the electronic Zeeman splitting and assume the initial state is diagonal (in the basis of \(H_{\gamma}'\)) and has its occupation (approximately) only in the GS, e.g.\ thermal states with cryogenic temperatures.
The dynamics of the reduced density matrix
\(\dot{\tilde{\rho}} = \frac{i}{\hbar} \comm{\tilde{\rho}}{H_{\mathrm{eff}}} + L_{\mathrm{eff}} \tilde{\rho} L_{\mathrm{eff}}^{\dag} - \frac{1}{2} \{ L_{\mathrm{eff}}^{\dag} L_{\mathrm{eff}}, \tilde{\rho} \}\)
simplify to the dynamics of the diagonal entries
\begin{align}
  \label{eq:rateEQa}
  \dot{\tilde{\rho}}_{\uparrow,m} = - \Gamma_{\uparrow,m \to \downarrow,m} \rho_{\uparrow,m} + \Gamma_{\downarrow,m-1 \to \uparrow,m} \rho_{\downarrow,m-1}, \\
  \label{eq:rateEQb}
  \dot{\tilde{\rho}}_{\downarrow,m} = - \Gamma_{\downarrow,m \to \uparrow,m+1} \rho_{\downarrow,m} + \Gamma_{\uparrow,m \to \downarrow,m} \rho_{\uparrow,m} .
\end{align}
If the electric field amplitudes are chosen such that the resonant Rabi frequencies are all equal $\Omega_r$ we can approximate all rates with
$\Gamma_{\mathrm{eff}} = 4 \Omega_r^2 / \Gamma_{\mathrm{tot}}$.
Additionally assuming that the population of all levels of the GS KD are equal, the analytic solution of the rate Eqs.~\eqref{eq:rateEQa}~and~\eqref{eq:rateEQb}
can be used to very compactly write the solution for the occupation of the final state (for the nuclear spin $I=7/2$ of vanadium)
\begin{align}
    \label{eq:avRate}
    \tilde{\rho}_{\downarrow,I} \approx 1 - \frac{e^{-\Gamma_{\mathrm{eff}} t}}{16} \sum_{k=0}^{14} \frac{15-k}{k!} \Gamma_{\mathrm{eff}}^k t^k
\end{align}

We emphasize that while these rates and the solution of the system are suited to estimate the timescale of the dynamics and to explain the final state,
in order to describe the full dynamics, a description involving the full time evolution is better suited as it is numerically feasible and contains several aspects neglected here for simplicity.

\end{widetext}

\bibliography{refs.bib}

\end{document}